\documentclass[10pt,journal]{IEEEtran}

\usepackage{amsmath}
\usepackage{amsfonts}
\usepackage{graphicx}
\usepackage{float}
\usepackage{tabularx}
\usepackage{amssymb}
\usepackage{pdflscape}
\usepackage{setspace}
\usepackage{color}
\usepackage{graphics}
\usepackage{subfig}
\usepackage{epsfig}
\usepackage[T1]{fontenc}
\usepackage{times}
\usepackage[hyphens]{url}
\usepackage{comment}
\usepackage{algorithmicx,algorithm}
\usepackage{bbm}
\usepackage{algpseudocode}
\usepackage{pifont}
\usepackage{cite}
\usepackage{multirow}
\usepackage{booktabs}

\title{{\small Accepted in the 58th North American Power Symposium, Houghton, Michigan 2026 (NAPS 2026)}\\Contingency Detection Integrated Model Predictive Control for Resilient Load Frequency Control 
\thanks{This work was supported in part by the National Science Foundation of USA under Grants ECCS-2146615, CNS-2231523, and the eCAT Center industry members.}
}

\author{ Erfan Mehdipour Abadi, Hamid Varmazyari, Sharaf K. Magableh, Caisheng Wang, Le Yi Wang, Feng Lin\\ \textit{Department of Electrical and Computer Engineering}\\ \textit{Wayne State University}\\ Detroit, MI, United States\\ \{erfan.abadi, varmazyari.h, sharaf.magableh, cwang, lywang, flin\}@wayne.edu 
\vspace{-0.75cm}}

\begin{document}

\maketitle

\begin{abstract}
Contingencies can alter power-system dynamics and introduce prediction mismatch in model predictive control (MPC)-based load frequency control (LFC). Although such events may be detected or cleared by protection systems, the corresponding post-contingency dynamic model may not be available to the MPC controller on the LFC time scale. This paper proposes a contingency detection-integrated MPC (CDI-MPC) framework that combines disturbance-aware contingency detection with predictive frequency regulation. Contingencies are modeled as stochastic discrete events of a stochastic hybrid system (SHS), and a disturbance-aware residual formulation is developed to jointly identify the active mode and estimate unknown disturbances. The detected mode is then used to update the MPC prediction model, reducing contingency-induced prediction mismatch under changing operating conditions. Simulation results demonstrate accurate contingency detection and substantial improvements in closed-loop LFC performance under multiple contingency scenarios and unknown disturbances.

\end{abstract}

\section{Introduction}
Load frequency control (LFC) is becoming increasingly important in modern power systems due to the growing integration of inverter-based resources (IBRs), electric vehicles (EVs), flexible loads, and large data-center demands~\cite{MASIKANA2024100605}. Compared with conventional systems, these resources introduce faster dynamics, larger fluctuations, and reduced inertia, making reliable frequency regulation increasingly challenging~\cite{10759610}. Meanwhile, the increasing complexity of power-system operation has heightened the need for detecting contingencies such as topology changes, line outages, and component failures, which alter the system dynamics~\cite{abadi2024distributed}. Although protection systems may detect these events, the corresponding post-contingency model is typically unavailable to the LFC controller on the frequency-control time scale.

Model predictive control (MPC) has been widely investigated as a
promising approach for LFC because its prediction-based structure can incorporate system dynamics, operational constraints, and finite-horizon performance objectives into the control problem~\cite{wadi2024load,rasolomampionona2024comprehensive,ersdal2015model}. Robust MPC-based LFC has addressed uncertainties, time-varying dynamics, communication delays, cyber attacks, and measurement integrity~\cite{ojaghi2017lmi,hu2023resilient}, while adaptive MPC methods update prediction models using inertia estimation, Kalman-filter-based estimation, or recursive identification~\cite{yang2020inertia,wang2024frequency,nagaosa2025adaptive,ayman2026adaptive}. However, these approaches either accommodate uncertainty without identifying the active operating mode or update continuous parameters rather than the post-contingency model. Consequently, prediction mismatch after contingencies can degrade frequency-regulation performance. This work therefore aims to provide the MPC controller with online awareness of the active system-wide dynamic mode.

In our previous work, contingencies were modeled as random discrete events within a stochastic hybrid system (SHS), where each contingency corresponds to a distinct predefined operating mode identified from measured input-output data~\cite{yuan2024contingency,11225525}. A probing-based framework was developed to guarantee distinguishability among candidate modes over a short detection window~\cite{wang2023joint}. However, extending this framework to closed-loop MPC-based LFC requires accounting for both the MPC-generated control input and unknown load disturbances during detection. Unlike the disturbance-free setting in~\cite{wang2023joint}, load disturbances are intrinsic to LFC and must be distinguished from contingency-induced dynamics.

In this paper, we develop a Contingency Detection-Integrated MPC (CDI-MPC) framework for resilient LFC under contingencies and unknown load disturbances. Contingencies are modeled as predefined operating modes of an SHS model, where each mode corresponds to a possible post-contingency dynamic model. Since the active mode is not assumed to be directly available to the MPC controller, contingency awareness is formulated as an online mode-identification problem. To prevent load disturbances from being misinterpreted as contingency-induced dynamics, the detection stage constructs disturbance-aware residuals by explicitly accounting for the applied control and probing input and jointly detecting the actual mode and estimating the disturbance over a short detection window. The detected mode and disturbance estimate are then embedded into the MPC prediction model, enabling mode-dependent prediction updates and reducing contingency-induced prediction mismatch in the closed-loop LFC response.

Main contributions of this paper are summarized as follows:
\begin{itemize}
\item A disturbance-aware extension of SHS-based contingency detection is developed for closed-loop LFC operation. The proposed detection formulation explicitly accounts for the applied control and probing input and jointly estimates the unknown initial state and load disturbance, preventing disturbance effects from being misinterpreted as contingency dynamics.

\item A CDI-MPC framework is proposed to incorporate the detected operating mode and disturbance estimate into the MPC prediction model. By replacing the nominal model with a mode-dependent prediction model, the proposed controller mitigates contingency-induced prediction mismatch and improves LFC performance under contingencies.
\end{itemize}

The remainder of this paper is organized as follows. Section~\ref{Sec-II} formulates the MPC-based LFC problem under contingencies and characterizes the resulting prediction mismatch. Section~\ref{Sec-III} presents the proposed CDI-MPC framework, including disturbance-aware mode detection and mode-dependent MPC prediction updating. Section~\ref{Sec-IV} provides simulation studies to evaluate detection accuracy and closed-loop frequency regulation performance under contingencies and load disturbances. Finally, Section~\ref{Sec-V} concludes the paper.

\section{Problem Formulation}\label{Sec-II}
This section formulates the MPC-based LFC problem under contingencies. First, a nominal MPC formulation is introduced. Then, contingencies are represented as operating modes of a randomly switched linear system (RSLS) within the SHS framework. Finally, the resulting prediction mismatch under contingency conditions is characterized.

\subsection{MPC-Based Load Frequency Control}
The LFC dynamics considered in this work describe the frequency response of the system under control actions from participating resources and load disturbances. Around a nominal operating point, these dynamics are linearized and represented in the continuous-time state-space form
\begin{align} \label{eq:nominal_lfc_ct}
\dot{x}(t) &= A_0^c x(t)+B_0^c u(t)+E_0^c w(t), \\  \notag
y(t) &= C_0 x(t),
\end{align}
where $x(t)$ is the system state vector, $u(t)$ is the LFC control input, $w(t)$ denotes the load disturbance, and $y(t)$ is the measured frequency-related output, such as frequency deviation or area control error. The subscript $0$ denotes the nominal operating condition, while the superscript $c$ indicates the continuous-time representation of the system matrices.

For MPC implementation, \eqref{eq:nominal_lfc_ct} is discretized with sampling time $T_s$ using a zero-order hold, yielding
\begin{align}
x(k+1) &= A_0x(k)+B_0u(k)+E_0w(k), \\ \notag
y(k) &= C_0x(k),
\label{eq:nominal_lfc_dt}
\end{align}
where $A_0$, $B_0$, and $E_0$ are the nominal discrete-time matrices obtained from the discretization.

Over a prediction horizon $N$, define the stacked state trajectory as
\begin{equation}
X_k^j=
\begin{bmatrix}
x(k+1)\\
x(k+2)\\
\vdots\\
x(k+N)
\end{bmatrix}.
\label{eq:mpc_stacked_state}
\end{equation}
Similarly, $U_k^j$ and $W_k^j$ denote the stacked input and disturbance
sequences associated with the $j$th operating mode, whose discrete-time
state-space matrices are denoted by $(A_j,B_j,E_j)$.

For the $j$th operating mode, the lifted prediction model assumes that
the operating mode remains unchanged over the prediction horizon, since
future mode transitions are unknown to the MPC controller. Accordingly,
the lifted prediction model is given by
\begin{equation}
 X_k^j=\Phi_jx(k)+\Gamma_jU_k^j+\Omega_jW_k^j,
\label{eq:generic_lifted_model}
\end{equation}
where $ X_k^j$ denotes the state trajectory predicted using $(A_j,B_j,E_j)$, and
\begin{equation}
\Phi_j \triangleq
\begin{bmatrix}
A_j^1\\
A_j^2\\
\vdots\\
A_j^N
\end{bmatrix},
\quad
\Gamma_j \triangleq
\begin{bmatrix}
B_j & 0 & \cdots & 0\\
A_jB_j & B_j & \cdots & 0\\
\vdots & \vdots & \ddots & \vdots\\
A_j^{N-1}B_j & A_j^{N-2}B_j & \cdots & B_j
\end{bmatrix},
\label{eq:Phi_Gamma_j}
\end{equation}
and
\begin{equation}
\Omega_j \triangleq
\begin{bmatrix}
E_j & 0 & \cdots & 0\\
A_jE_j & E_j & \cdots & 0\\
\vdots & \vdots & \ddots & \vdots\\
A_j^{N-1}E_j & A_j^{N-2}E_j & \cdots & E_j
\end{bmatrix}.
\label{eq:Omega_j}
\end{equation}

For the nominal LFC model, the lifted matrices are obtained by using $A_0$, $B_0$, and $E_0$ in \eqref{eq:Phi_Gamma_j}--\eqref{eq:Omega_j}. Since the future load disturbance is not available to the nominal controller, the nominal MPC prediction is constructed assuming $W_k^0 = \mathbf{0}$. Accordingly, the disturbance-free nominal prediction is given by
\begin{equation}
\bar X_k^0=\Phi_0x(k)+\Gamma_0U_k^0,
\label{eq:nominal_lifted_model}
\end{equation}
where $\bar X_k^0$ denotes the state trajectory predicted by the nominal model in the absence of disturbance forecasting.

Following standard MPC-based LFC formulations~\cite{ersdal2015model,ojaghi2017lmi}, the nominal controller computes the input sequence by solving a finite-horizon quadratic programming problem that penalizes predicted state deviations from the
nominal operating point and control effort. Using the disturbance-free nominal prediction in \eqref{eq:nominal_lifted_model}, the nominal MPC problem is formulated as
\begin{subequations}
\label{eq:nominal_mpc_qp}
\begin{align}
\min_{U_k} \quad
& J_0(k)
=
(\bar X_k^0)^T\bar Q\bar X_k^0
+
U_k^T\bar R U_k
\label{eq:nominal_mpc_qp_obj}
\\
\text{s.t.} \quad
& \bar X_k^0=\Phi_0x(k)+\Gamma_0U_k,
\label{eq:nominal_mpc_qp_dyn}
\\
& U_{\min}\leq U_k\leq U_{\max},
\label{eq:nominal_mpc_qp_input}
\\
& X_{\min}\leq \bar X_k^0\leq X_{\max}.
\label{eq:nominal_mpc_qp_state}
\end{align}
\end{subequations} Here, $\bar Q \succeq 0$ and $\bar R \succ 0$ are the horizon state-
and input-weighting matrices, respectively, where $\bar Q$ is
positive semi-definite and $\bar R$ is positive definite. The matrix $\bar Q$ determines the relative importance of frequency regulation performance, while $\bar R$ penalizes aggressive control actions and reflects the desired use of participating regulation resources. Constraint \eqref{eq:nominal_mpc_qp_dyn} enforces the nominal lifted prediction model, \eqref{eq:nominal_mpc_qp_input} represents input limitations such as resource capacity and actuation bounds, and \eqref{eq:nominal_mpc_qp_state} imposes admissible bounds on the predicted state trajectory, including frequency-security requirements.
After solving \eqref{eq:nominal_mpc_qp}, the first element of the optimal input sequence is applied to the system, and the optimization is repeated at the next sampling instant using updated measurements; however, this baseline formulation relies on the nominal prediction model and does not account for contingencies in the active system dynamics.

\subsection{Contingencies as RSLS Modes}\label{II-B}

Following the RSLS modeling framework in
\cite{wang2023joint,yuan2024contingency, 11225525}, contingencies are
represented as random discrete events that change the active
state-space model of the system. This representation is particularly suitable for LFC because the system is linearized around each operating condition. Consequently, the nominal operating condition and each contingency scenario are represented by distinct linear subsystems, with each subsystem corresponding to a unique operating mode of the SHS. When a contingency occurs, the system transitions from the current operating mode to the corresponding post-contingency mode, resulting in a switch between the associated linear state-space models. In the LFC setting considered here, a contingency refers to an unobserved change in the dynamic model
that affects the frequency response before the event is explicitly
identified by the operator or controller.

Let $\alpha_k\in\mathcal{S}=\{0,1,\ldots,M\}$ denote the active
operating mode at time $k$, where $\alpha_k=0$ corresponds to the
nominal condition and $\alpha_k=i$, $i=1,\ldots,M$, corresponds to the
$i$th predefined contingency mode. The contingency-dependent dynamics
are represented as
\begin{align}
x(k+1)
&=
A_{\alpha_k}x(k)
+
B_{\alpha_k}u(k)
+
E_{\alpha_k}w(k), \notag\\
y(k)
&=
C_{\alpha_k}x(k).
\label{eq:rsls_model}
\end{align}
For a fixed mode $\alpha_k=i$, the system evolves according to the
linear subsystem $(A_i,B_i,E_i,C_i)$. Thus, contingencies are
modeled as switches among a finite set of candidate dynamic modes.
The mode set, $\mathcal{S}$ is assumed to be known, while the switching sequence
$\alpha_k$ and the switching time are not directly available to the controller, and must be inferred from measured input-output data over the detection window. The candidate mode set $\mathcal{S}$ is constructed offline from
the planning stage of contingency studies, in which credible and operationally
significant contingencies are identified and represented by their
corresponding linearized models. Accordingly, each candidate mode
corresponds to a predefined contingency scenario. This work assumes that
the relevant operating contingencies are represented in
$\mathcal{S}$. Extending the proposed framework to broader contingency
classes or clustered contingency models is left for future work.

\subsection{Prediction Mismatch under Contingencies}

The nominal MPC problem in \eqref{eq:nominal_mpc_qp} is solved using
the disturbance-free nominal prediction $\bar X_k^0$ in
\eqref{eq:nominal_lifted_model}. If the system remains in the nominal
mode, the actual lifted state trajectory over the horizon is
\begin{equation}
X_k^0=\Phi_0x(k)+\Gamma_0U_k+\Omega_0W_k .
\label{eq:actual_nominal_lifted}
\end{equation}
Therefore, the prediction error under the nominal model is
\begin{equation}
\Delta X_k^{0}
\triangleq
X_k^0-\bar X_k^0
=
\Omega_0W_k .
\label{eq:nominal_disturbance_mismatch}
\end{equation}
This term represents the effect of the load disturbance, which is the
standard regulation task addressed by LFC through receding-horizon
feedback.

In contrast, when a contingency $i\neq 0$ occurs, the actual
state trajectory for the active mode $i$ over the prediction horizon is
\begin{equation}
X_k^i=\Phi_i x(k)+\Gamma_i U_k+\Omega_i W_k,
\label{eq:actual_contingency_lifted}
\end{equation}
where $(\Phi_i,\Gamma_i,\Omega_i)$ are constructed from
$(A_i,B_i,E_i)$ using \eqref{eq:Phi_Gamma_j}--\eqref{eq:Omega_j}. The
prediction mismatch between the active contingency model and the disturbance-free nominal model used by MPC is then
\begin{align}
\Delta X_k^{i|0}
&\triangleq
X_k^i-\bar X_k^0 \notag\\
&=
(\Phi_i-\Phi_0)x(k)
+
(\Gamma_i-\Gamma_0)U_k
+
\Omega_iW_k .
\label{eq:contingency_prediction_mismatch}
\end{align}
Unlike \eqref{eq:nominal_disturbance_mismatch}, the mismatch in
\eqref{eq:contingency_prediction_mismatch} contains not only the effect
of load disturbance, but also the mismatch between the nominal and active
mode-dependent state and input prediction operators. Hence, even in the
absence of load disturbance, an unidentified contingency can produce a
nonzero prediction mismatch if $(\Phi_i,\Gamma_i)\neq(\Phi_0,\Gamma_0)$.

As a result, the equality constraint in \eqref{eq:nominal_mpc_qp_dyn} and
the state trajectory penalized in \eqref{eq:nominal_mpc_qp_obj} may no
longer represent the actual system evolution. Therefore, maintaining
accurate predictive regulation requires identifying the active operating
mode and adapting the MPC prediction model accordingly. Moreover, since
the active mode is unknown to the controller, the initial condition associated with each
candidate model cannot be assumed known during detection. This additional
uncertainty, together with the unknown load disturbance, motivates the
disturbance-aware contingency detection and prediction-model updating
framework developed in the next section.

\section{Contingency Detection-Integrated MPC}\label{Sec-III}
This section presents the proposed CDI-MPC framework. Building on the probing-based RSLS detection framework, it extends closed-loop LFC by accounting for the applied control input and unknown load disturbance during mode identification. The detected mode and disturbance estimate are then incorporated into the MPC prediction model.

\subsection{RSLS-Based Detection Framework Preliminaries}

The RSLS-based detection framework adopted from
\cite{wang2023joint,yuan2024contingency} operates in a time-division
structure. In this paper, this structure is embedded into a closed-loop MPC-based LFC
by partitioning the closed-loop operation into repeated update intervals.
Let $L$ denote the number of samples in each update interval and let
$k_\ell=\ell L$ denote the starting instant of the $\ell$th interval.
The first $N_d$ samples, with $N_d\ll L$, form the detection window
\begin{equation}
\mathcal{D}_\ell
=
\{k_\ell,\ldots,k_\ell+N_d-1\}.
\label{eq:detection_window}
\end{equation}
During this window, the LFC control input remains active while a probing signal is superimposed on the input to identify the active mode dynamics. This probing signal is adopted from the input-design procedure in
\cite{wang2023joint}, which is used to guarantee distinguishability among
candidate RSLS modes when passive measurements alone are insufficient due to similarities in the behaviors of different contingencies.

At the end of $\mathcal{D}_\ell$, the data collected over the detection window are used to identify the active mode. The remaining samples of the same update interval form the regulation phase
\begin{equation}
\mathcal{R}_\ell
=
\{k_\ell+N_d,\ldots,k_\ell+L-1\}.
\label{eq:regulation_window}
\end{equation}
During $\mathcal{R}_\ell$, the probing signal is removed, and the MPC controller performs frequency regulation using the most recently updated
prediction model.

Accordingly, the applied input during the detection window is
\begin{equation}
u(k)=u_c(k)+u_p(k), \qquad k\in\mathcal{D}_\ell,
\label{eq:control_probe_input}
\end{equation}
where $u_c(k)$ is the LFC control input generated by MPC and $u_p(k)$ is
the probing signal. During the regulation phase, $u_p(k)=0$.

The choice of $L$ and $N_d$ creates a delay-performance tradeoff. Since contingencies are unknown, events that arise during the regulation phase are detected only at the next update instant. Therefore, reducing $L$ decreases the maximum detection delay and enables faster controller adaptation, but increases the probing duty ratio $N_d/L$, resulting in more frequent probing injections and potentially larger temporary frequency deviations. On the other hand, the minimum required value of $N_d$ can be selected according to the theoretical result in~\cite{wang2023joint}, which guarantees distinguishability among the candidate operating modes. Increasing $N_d$ generally improves estimation accuracy by incorporating more measurements, but also delays the controller update following a contingency. Consequently, $L$ and $N_d$ should be selected jointly to balance detection speed, estimation accuracy, and closed-loop LFC performance.

During the detection window,  $\mathcal{D}_\ell$, the measured output sequence is collected as
\begin{equation}
Y_\ell =
\begin{bmatrix}
y(k_\ell+1)\\
y(k_\ell+2)\\
\vdots\\
y(k_\ell+N_d)
\end{bmatrix}.
\label{eq:detection_output_stack}
\end{equation}
The probing-based RSLS detection principle in
\cite{wang2023joint,yuan2024contingency} compares this measured response, $Y_\ell$,
with the responses reconstructed from the candidate modes. Let
$\widehat{Y}_{\ell,i}$ denote the output sequence reconstructed over
$\mathcal{D}_\ell$ under candidate mode $i\in\mathcal{S}$. The mode-dependent residual is then defined as
\begin{equation}
r_i(\ell)
=
\left\|
Y_\ell-\widehat{Y}_{\ell,i}
\right\|_2^2,
\qquad \forall i\in\mathcal{S},
\label{eq:generic_detection_residual}
\end{equation}
and the active mode is detected by the minimum residual
\begin{equation}
\hat{\alpha}_\ell
=
\arg\min_{i\in\mathcal{S}} r_i(\ell).
\label{eq:generic_mode_selection}
\end{equation}

However, in the probing-based detection setting used in
\cite{wang2023joint,yuan2024contingency}, the residual in
\eqref{eq:generic_detection_residual} is constructed for an open-loop,
disturbance-free system response, where the probing signal is the main
known input used for mode detection, and the initial condition is the only
unknown parameter. By contrast, in the closed-loop LFC setting considered
here, the measured response over $\mathcal{D}_\ell$ is affected by both
the MPC regulation input and the probing signal, as well as the unknown
load disturbances. Moreover, the initial condition associated with each candidate mode is also unknown during detection. Therefore, the main task
is to construct $\widehat{Y}_{\ell,i}$ by explicitly accounting for the
known applied input while jointly estimating the candidate initial
condition and load disturbance. This disturbance-aware reconstruction is
developed next.
\vspace{-0.2cm}
\subsection{Disturbance-Aware Contingency Detection}

During the detection window $\mathcal{D}_\ell$, the measured output
sequence $Y_\ell$ is defined in \eqref{eq:detection_output_stack}, and
let $U_\ell$ denote the stacked applied input sequence over the same
window. Since $\mathcal{D}_\ell$ is short, the load disturbance is
approximated as constant over the detection window,
\begin{equation}
w(k)\approx d_\ell,\qquad k\in\mathcal{D}_\ell,
\label{eq:constant_disturbance_detection}
\end{equation}
where $d_\ell$ is the unknown constant disturbance over the detection window.

For each possible mode $i\in\mathcal{S}$, the lifted prediction model
in \eqref{eq:generic_lifted_model} is applied over the detection horizon
$N_d$ using the corresponding state-space matrices $(A_i,B_i,E_i)$.
Thus, the candidate state trajectory over $\mathcal{D}_\ell$ is written as
\begin{equation}
X_{\ell}^{i}
=
\Phi_i x(k_\ell)+\Gamma_i U_\ell+\Omega_i W_\ell .
\label{eq:detection_state_lifted}
\end{equation}
where $\Phi_i$, $\Gamma_i$, and $\Omega_i$ are obtained from
\eqref{eq:Phi_Gamma_j} and \eqref{eq:Omega_j} by setting the horizon
length to $N_d$. Under the constant-disturbance approximation in
\eqref{eq:constant_disturbance_detection}, the disturbance sequence is
given by
$W_\ell=\mathbf{1}_{N_d}\otimes d_\ell$ .

Let
$\mathcal{C}_i \triangleq I_{N_d}\otimes C_i$. Since the applied input sequence $U_\ell$ is known, its contribution to
the measured output can be removed for each candidate mode. The net
output is computed as
\begin{equation}
Y_{\ell,i}^{\mathrm{net}}
=
Y_\ell-\mathcal{C}_i\Gamma_iU_\ell .
\label{eq:ynet}
\end{equation}

After the known input contribution is removed, the remaining output response is attributed to the unknown initial condition and the unknown
load disturbance. Using \eqref{eq:detection_state_lifted} and the
constant-disturbance approximation
in \eqref{eq:constant_disturbance_detection}, the net output can be written as
\begin{equation}
Y_{\ell,i}^{\mathrm{net}}
=
\Lambda_i\theta_\ell+\eta_\ell,
\label{eq:ynet_regression}
\end{equation}
where $\eta_\ell$ denotes measurement noise,
\begin{equation}
\theta_\ell=
\begin{bmatrix}
x(k_\ell)\\
d_\ell
\end{bmatrix},
\qquad
\Lambda_i
=
\mathcal{C}_i
\begin{bmatrix}
\Phi_i & \Omega_i
\end{bmatrix}.
\label{eq:theta_lambda}
\end{equation}

For each candidate mode $i\in\mathcal{S}$, the unknown initial condition
and constant disturbance are estimated by fitting the net output response
in \eqref{eq:ynet}. Specifically,
\begin{equation}
\hat{\theta}_{\ell,i}
=
\arg\min_{\theta}
\left\|
Y_{\ell,i}^{\mathrm{net}}-\Lambda_i\theta
\right\|_2^2 ,
\label{eq:theta_estimation}
\end{equation}
where $\theta$ has the same structure as $\theta_\ell$ in
\eqref{eq:theta_lambda}. Assuming that $\Lambda_i$ has full column rank, the least-squares solution is
\begin{equation}
\hat{\theta}_{\ell,i}
=
(\Lambda_i^T\Lambda_i)^{-1}\Lambda_i^T
Y_{\ell,i}^{\mathrm{net}} .
\label{eq:theta_estimation_closed_form}
\end{equation}

If the full column rank condition is not satisfied,
the observable-subspace decomposition proposed in~\cite{wang2023joint} can be
employed to perform the estimation using only the observable component of
the system. Accordingly, the generic residual in \eqref{eq:generic_detection_residual}
is evaluated in a disturbance-aware form as
\begin{equation}
r_i(\ell)
=
\left\|
Y_{\ell,i}^{\mathrm{net}}
-
\Lambda_i\hat{\theta}_{\ell,i}
\right\|_2^2 .
\label{eq:disturbance_aware_residual}
\end{equation}

Thus, the detected contingency mode is obtained by \eqref{eq:generic_mode_selection} and the associated disturbance is
estimated as
\begin{equation}
\hat{d}_\ell
=
\begin{bmatrix}
0 & I
\end{bmatrix}
\hat{\theta}_{\ell,\hat{\alpha}_\ell}.
\label{eq:detected_mode_disturbance}
\end{equation}
These detections and estimation results are used to update the MPC prediction model during the regulation interval
$\mathcal{R}_\ell$. Thus, the controller replaces the nominal prediction
model with the model associated with the detected mode, and incorporates
the estimated disturbance, reducing the prediction mismatch caused by
contingencies and load variations.
\vspace{-0.7cm}
\subsection{CDI-MPC Formulation and Implementation}
\vspace{-0.2cm}
After contingency detection $\hat{\alpha}_\ell$ and load disturbance estimation $\hat d_\ell$, these parameters are used
to update the prediction model in the MPC problem. During the regulation phase
$\mathcal{R}_\ell$, the CDI-MPC problem is formulated as
\begin{subequations}
\label{eq:cdi_mpc_qp}
\begin{align}
\min_{U_k} \quad
& J_{C}(k)
=
(\hat X_k^{\hat{\alpha}_\ell})^T\bar Q
\hat X_k^{\hat{\alpha}_\ell}
+
U_k^T\bar R U_k
\label{eq:cdi_mpc_qp_obj}
\\
\text{s.t.} \quad
& \hat X_k^{\hat{\alpha}_\ell}
=
\Phi_{\hat{\alpha}_\ell}x(k)
+
\Gamma_{\hat{\alpha}_\ell}U_k
+
\Omega_{\hat{\alpha}_\ell}
W_\ell,
\label{eq:cdi_mpc_qp_dyn}
\\
& U_{\min}\leq U_k\leq U_{\max},
\label{eq:cdi_mpc_qp_input}
\\
& X_{\min}\leq \hat X_k^{\hat{\alpha}_\ell}\leq X_{\max}.
\label{eq:cdi_mpc_qp_state}
\end{align}
\end{subequations}
In \eqref{eq:cdi_mpc_qp}, $\hat X_k^{\hat{\alpha}_\ell}$ is the predicted
state trajectory generated by the detected contingency mode over the MPC
horizon. It is obtained from the lifted model in
\eqref{eq:generic_lifted_model} by setting $j=\hat{\alpha}_\ell$ and
using the estimated disturbance $\hat d_\ell$. Therefore, the objective in
\eqref{eq:cdi_mpc_qp_obj} penalizes the updated predicted state trajectory
and control effort, while the constraint in \eqref{eq:cdi_mpc_qp_state}
imposes the state and frequency limits on the same updated prediction.
Compared with the nominal MPC problem in \eqref{eq:nominal_mpc_qp}, the
CDI-MPC problem replaces the nominal prediction model with the
detected mode model and incorporates the estimated load disturbance.

Algorithm~\ref{alg:cdi_mpc} summarizes the online implementation of the proposed CDI-MPC framework. The computational burden is limited because
the candidate mode models, lifted prediction matrices, regression matrices,
and probing-input design are computed offline. During operation, the detector
is executed only once per update interval, jointly estimating the load
disturbance and identifying the active mode by evaluating short-window
residuals over the candidate modes. The MPC quadratic program is then solved
at each sampling instant using the detected mode. Therefore, compared with
nominal MPC, the additional computational cost of CDI-MPC is primarily the
intermittent residual-based mode-detection step.

\begin{algorithm}[t!]
\caption{Online CDI-MPC Implementation}
\label{alg:cdi_mpc}
\begin{algorithmic}[1]
\State Initialize $\hat{\alpha}=0$ and $\hat d=0$.
\For{$k=0,1,\ldots$}
    \State Set $k_L=\mathrm{mod}(k,L)$.
    \If{$k_L=N_d$}
        \State Detect the contingency mode and estimate the load disturbance
        using \eqref{eq:theta_estimation}-\eqref{eq:detected_mode_disturbance}.
        \State Update $\hat{\alpha}$ and $\hat d$.
    \EndIf
    \State Solve \eqref{eq:cdi_mpc_qp} using the current $\hat{\alpha}$ and $\hat d$.
    \If{$k_L<N_d$}
        \State Apply $u(k)=u_c(k)+u_p(k)$ and collect input-output data.
    \Else
        \State Apply $u(k)=u_c(k)$.
    \EndIf
\EndFor
\end{algorithmic}
\end{algorithm}

\section{Case Study}\label{Sec-IV}

\begin{figure}[t]
\centering
\includegraphics[
width=\columnwidth,
trim={0 1.4cm 0 1cm},
clip
]{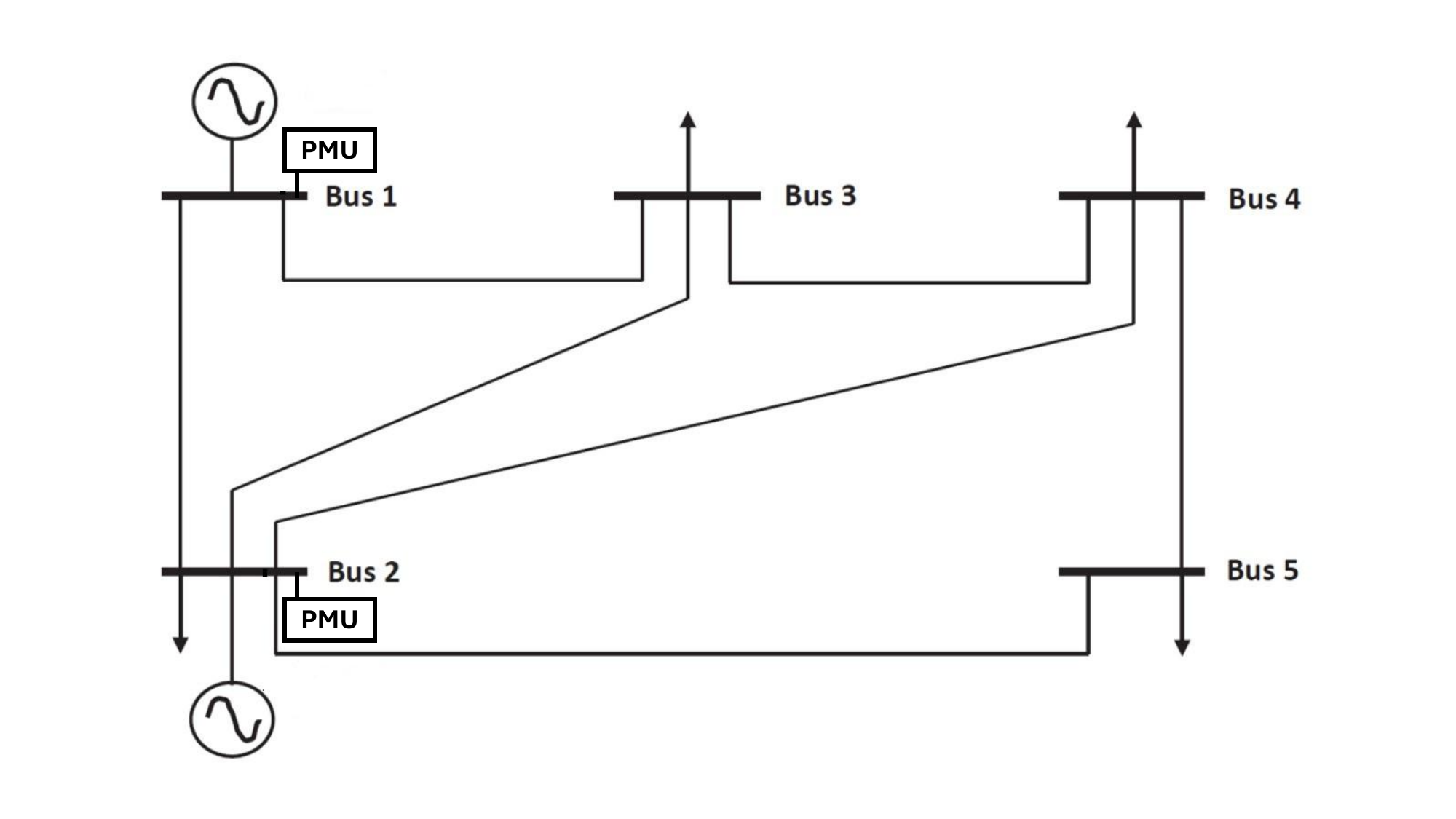}
\caption{IEEE 5-bus system.}
\label{fig:test_system}
\vspace{-0.75cm}
\end{figure}
The proposed CDI-MPC framework is evaluated using the 5-bus transmission network shown in Fig.~\ref{fig:test_system}. The network includes two generator buses, Buses~1 and~2, which are modeled as dynamic states of a reduced-order LFC system based on synchronous-generator swing equations. The remaining buses represent non-dynamic load buses whose influence on system dynamics is captured through algebraic network equations. The reduced-order LFC model is described by the state vector $x=[\delta_1,\omega_1,\delta_2,\omega_2]^T$, control input $u=[u_{1},u_{2}]^T$, and load disturbance input $w=[w_{1},w_{2}]^T$. Here, $u_1$ and $u_2$ denote the reference power commands applied to the generating units at Buses~1 and~2, respectively, while $w_1$ and $w_2$ represent the load disturbances acting at the corresponding buses. In the net output model of~(\ref{eq:ynet_regression}), the additive noise term $\eta_{\ell}$ is modeled as zero-mean Gaussian noise. The frequency- and rotor-angle measurement noise standard deviations are chosen as $\sigma_f = 10^{-4}\,\mathrm{Hz}$ and $\sigma_{\delta} = 10^{-4}\,\mathrm{rad}$, respectively. In general, reliable classification requires the noise-induced residual perturbation to remain sufficiently small relative to the residual margin between the actual mode and the nearest competing mode. Under higher measurement-noise levels or larger parameter uncertainties, increased probing energy, longer detection windows, or adaptive state and disturbance estimation may be required; their systematic design is left for future work. The system parameters and the system state-space model are introduced in \cite{yuan2024contingency}.

Contingencies are modeled through variations in the impedance of transmission line 1--3. Four candidate operating conditions are considered and are represented by four operating modes: Mode~0 for nominal operation with $Z_{13}=0.06$ p.u., Mode~1 for mild line damage with $Z_{13}=0.10$ p.u., Mode~2 for severe line damage with $Z_{13}=0.60$ p.u., and Mode~3 for line outage with $Z_{13}=1000$ p.u. These operating conditions define the RSLS mode set used by the detector and the CDI-MPC controller. The simulation horizon is $T_{\mathrm{sim}}=120$ s with sampling period
$T_s=0.1$ s. The contingency detection module is executed every
$L=10$ s. Within each update interval, the first $0.3$ s, corresponding to three measurement samples, is used as the detection interval
$\mathcal{D}_\ell$, and the remaining $9.7$ s forms the regulation
interval $\mathcal{R}_\ell$. The short detection interval demonstrates
that the proposed disturbance-aware RSLS detector can identify the
active operating mode and estimate the unknown disturbances using only
a small amount of measurement data, thereby minimizing the time spent
in detection and maximizing the time available for optimal LFC.

The MPC prediction and control horizons are selected as
$N_p=N_c=30$, corresponding to a $3$ s prediction horizon. The state
and input weighting matrices are chosen as
\[
Q=\mathrm{diag}(10,1000,10,1000), \qquad
R=10^{-1}I_2 .
\]
The larger weights assigned to the frequency states reflect the primary objective of LFC, namely, minimizing frequency deviations at the two
generator buses, while the smaller weights on the rotor-angle states
provide mild damping of angle deviations. The input penalty matrix is
selected to prevent excessive control effort while maintaining effective
frequency regulation performance.

The control inputs are constrained by
\[
-0.25\leq u_{1}\leq 0.25,\qquad
-0.15\leq u_{2}\leq 0.15,
\]
and the frequency deviations are limited to $\pm 0.5$ Hz. This bound is imposed as a conservative transient security constraint rather than a steady-state frequency-regulation requirement, consistent with commonly adopted LFC operating practices~\cite{SHARMA2024113609}. In the reported simulations, the resulting frequency deviations remain substantially below this bound. During the
detection interval, a sinusoidal probing signal
\[
u_p(t)=A_p\sin(\omega_p t)
\]
is injected through the first control channel, $u_1$, where
$A_p=0.02$ and $\omega_p=2\pi(0.8)$ rad/s. The probing signal is
designed according to the contingency-identification criteria proposed
in \cite{wang2023joint}. Since the probing amplitude is small
relative to the admissible control range, it provides sufficient
excitation for mode identification while introducing only a limited and
temporary impact on frequency regulation performance.

\begin{figure}[t]
\centering
\includegraphics[
width=0.85\columnwidth,
trim={0 6cm 0 6.5cm},
clip
]{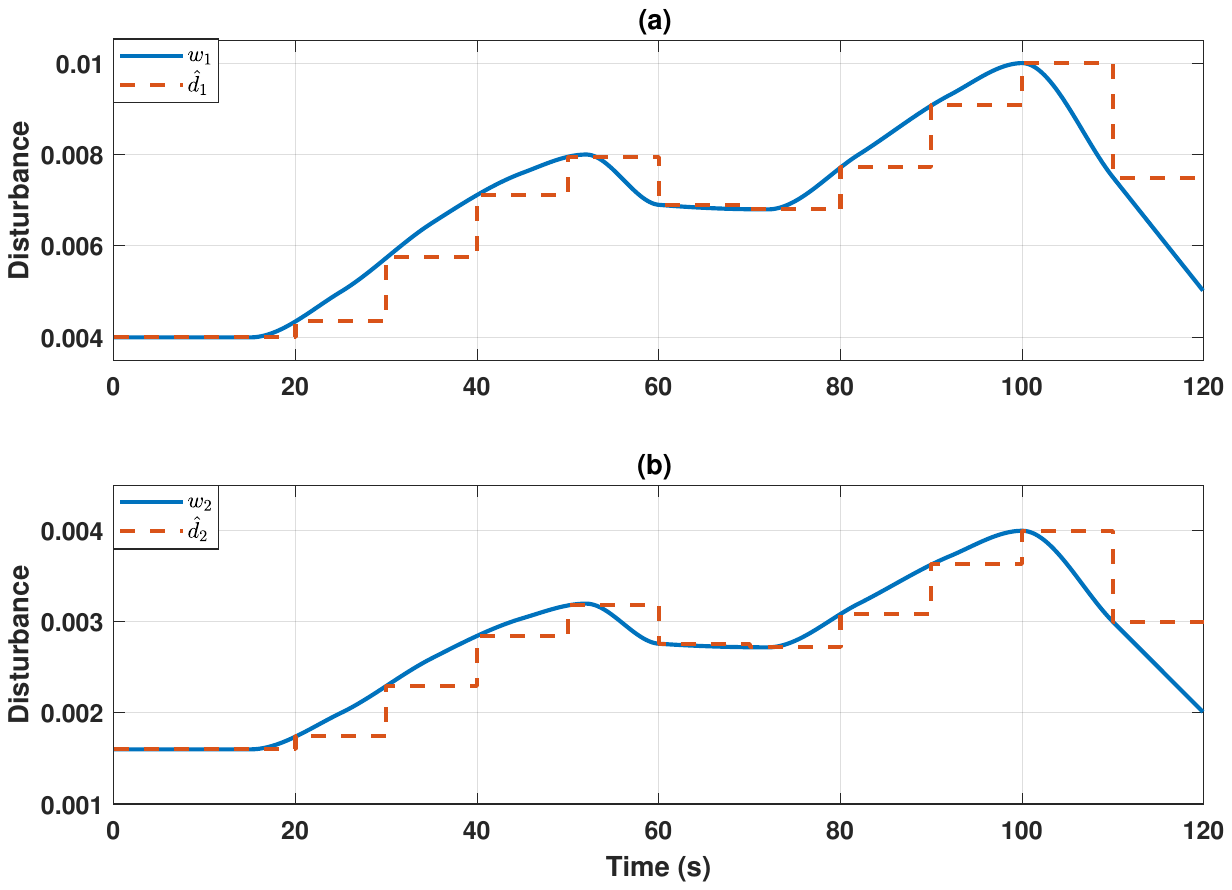}
\caption{Actual load disturbances, $w$, and estimated disturbance, $\hat{d}$ values for (a) Load~1 and (b) Load~2.}
\label{fig_load_estimation}
\end{figure}

Time-varying load disturbances are applied to both load buses, as shown in Fig.~2. Since the load profile varies slowly compared to the detection interval duration, the disturbance is assumed constant within each detection interval and is applied to the CDI-MPC framework during the subsequent regulation phase. Accordingly, the detector computes piecewise-constant estimates of the load disturbances, denoted by $\hat d_1$ and $\hat d_2$ for Loads~1 and~2, respectively, which closely track the underlying disturbance profiles. 

\begin{figure}[t]
\centering
\includegraphics[
width=0.85\columnwidth,
trim={0cm 10.5cm 0cm 10.75cm},
clip
]{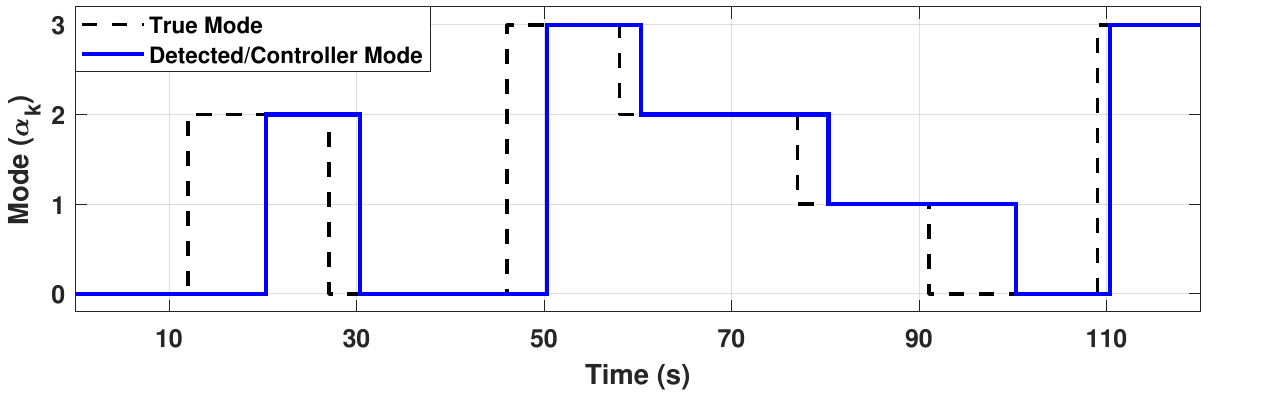}
\caption{Actual operating mode and detected controller mode under the proposed CDI-MPC framework.}
\label{fig:mode_detection}
\vspace{-0.5cm}
\end{figure}

To evaluate the mode-identification performance of the proposed disturbance-aware RSLS detector, Fig.~\ref{fig:mode_detection} compares the detected controller mode with the actual system mode. Since the contingency events are introduced at arbitrary instants and are not synchronized with the detection-update schedule, transitions occurring during a regulation interval are detected at the next detection interval. This explains the short delays observed between the actual and detected mode transitions in Fig.~\ref{fig:mode_detection}. The detection accuracy is defined as the percentage of scheduled detection updates for which the detected operating mode matches the true operating mode. Under this definition, the detector achieves a detection accuracy of $100\%$ over the 12 detection updates performed during the 120-s simulation, which include seven operating-mode transitions. Since mode estimation is performed only at the scheduled detection-update instants, the short delay between a contingency occurrence and the subsequent detection update is not counted as a misclassification. Instead, its effect is reflected in the closed-loop frequency-regulation performance metrics presented next.

\begin{figure}[t]
\centering
\includegraphics[
width=\columnwidth,
trim={0 5.5cm 0 5.5cm},
clip
]{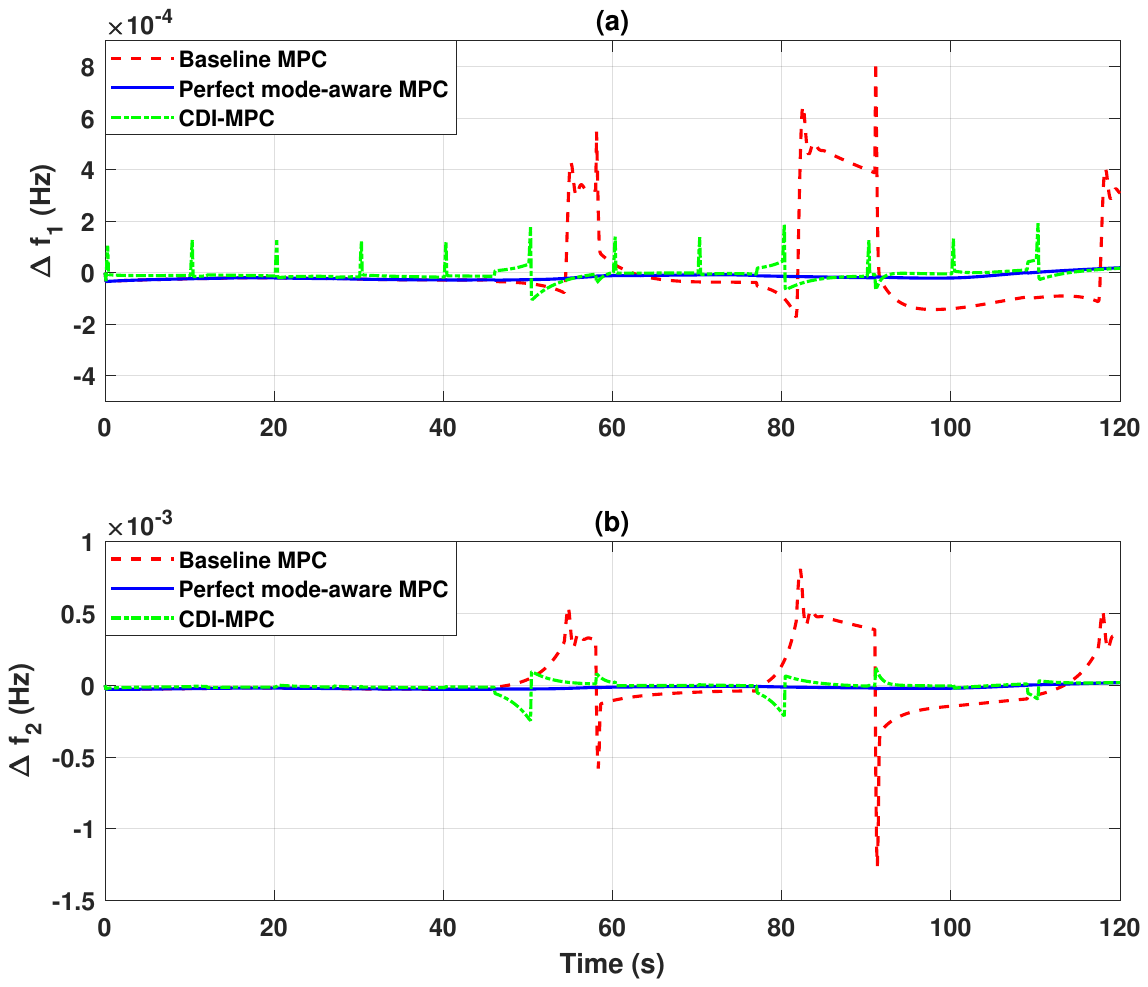}
\caption{Frequency deviations of (a) Generator~1 and (b) Generator~2 under baseline MPC, perfect mode-aware MPC, and the proposed CDI-MPC framework.}
\label{fig:frequency_response}
\vspace{-0.3cm}
\end{figure}

Having established the effectiveness of the proposed disturbance-aware
mode-detection framework, we next evaluate its impact on closed-loop LFC
performance. Fig.~\ref{fig:frequency_response} compares three control
strategies: baseline MPC, perfect mode-aware MPC, and the proposed
CDI-MPC. The baseline MPC uses a fixed nominal prediction model,
whereas the perfect mode-aware MPC assumes knowledge of the true
operating mode and serves as an ideal benchmark. CDI-MPC updates its
prediction model using the detected mode.

Figs.~\ref{fig:frequency_response}(a) and
\ref{fig:frequency_response}(b) compare the frequency deviations of
Generators~1 and~2, respectively. All three controllers regulate the
load variations and maintain the frequency within the admissible range.
Following contingency-induced mode transitions, the baseline MPC
exhibits larger frequency excursions because its nominal prediction
model no longer matches the system dynamics. In contrast, the perfect
mode-aware MPC continuously uses the correct model and is therefore much
less affected. The CDI-MPC response reflects two competing effects:
mode-dependent prediction updates reduce contingency-induced prediction
mismatch, whereas probing introduces small temporary deviations,
primarily in Generator~1 because the probing input is applied through
the first control channel.

\begin{table}[t]
\caption{Frequency-regulation performance comparison.}
\label{tab:integral_metrics}
\centering
\footnotesize
\setlength{\tabcolsep}{4pt}
\begin{tabular}{lcccc}
\hline
Metric & Baseline & Perfect & CDI-MPC & Improvement (\%)\\
\hline

IAE ($\Delta f_1$)
& 0.0117
& $2.18\times10^{-5}$
& 0.0019
& 83.5 \\

IAE ($\Delta f_2$)
& 0.0141
& $2.09\times10^{-5}$
& 0.0025
& 81.9 \\

ITAE ($\Delta f_1$)
& 0.949
& 0.0013
& 0.113
& 88.1 \\

ITAE ($\Delta f_2$)
& 1.131
& 0.0013
& 0.156
& 86.2 \\

\hline
\end{tabular}
\vspace{-0.3cm}
\end{table}

To quantify the temporary impact of the probing signal on
frequency-regulation performance, a contingency-free
simulation was performed over the interval $0$--$10$~s with and without
probing. The frequency-regulation performance is evaluated using the
Integral Absolute Error (IAE) and Integral Time-weighted Absolute Error
(ITAE), which measure the total magnitude and persistence of frequency
deviations, respectively. The maximum probe-induced frequency deviations
were $1.39\times10^{-4}\,\mathrm{Hz}$ and
$7.63\times10^{-6}\,\mathrm{Hz}$ for generators~1 and~2,
respectively. Over the same interval, the IAE of generator~1 increased
from $2.79\times10^{-4}$ to $3.00\times10^{-4}$, while the
corresponding ITAE increased from $1.32\times10^{-3}$ to
$1.36\times10^{-3}$. These results confirm that the probing signal
introduces only a small temporary degradation in
frequency-regulation performance while providing the excitation required
for reliable mode identification. Table~\ref{tab:integral_metrics} quantifies the overall
frequency-regulation performance of the three controllers using the IAE
and ITAE. CDI-MPC reduces the IAE and ITAE by approximately $84\%$ and $88\%$ for $\Delta f_1$, and by approximately $82\%$ and $86\%$ for $\Delta f_2$. These results show that the performance gain obtained from contingency identification and model updating outweighs the temporary probing effect. As expected, the perfect mode-aware MPC gives the best performance because it directly uses the true operating mode. Nevertheless, CDI-MPC remains much closer to this ideal benchmark than to the baseline MPC, demonstrating that the proposed framework can recover the operating-mode information needed for adaptive MPC-based LFC without direct access to the true contingency mode.

\section{Conclusion}\label{Sec-V}

This paper proposed a Contingency Detection-Integrated MPC (CDI-MPC)
framework for resilient load frequency control under contingencies and
unknown load disturbances. A disturbance-aware residual formulation was
developed to jointly identify the active operating mode and estimate the
load disturbance, and the resulting estimates were incorporated into the
MPC prediction model to mitigate contingency-induced prediction
mismatch.

Simulation results demonstrated accurate mode identification and
significant improvements in closed-loop frequency-regulation performance
over nominal MPC, with performance approaching the ideal mode-aware
benchmark. These results demonstrate the effectiveness of integrating
contingency detection with predictive control for resilient LFC.

\bibliographystyle{IEEEtran}
\bibliography{references}

\end{document}